\documentstyle[twoside,fleqn,psfig,espcrc2]{article}

\newcommand{\AmS}{{\protect\the\textfont2
  A\kern-.1667em\lower.5ex\hbox{M}\kern-.125emS}}
\newcommand{\be}{\begin{equation}}
\newcommand{\ee}{\end{equation}}
\newcommand{\bea}{\begin{eqnarray}}
\newcommand{\eea}{\end{eqnarray}}
\def\href#1#2{#2} 
\title{Electromagnetic dipole radiation of oscillating D-branes }

\author{G.K.Savvidy\address{National Research Center Demokritos,\\ 
        Ag. Paraskevi, GR-15310 Athens,Greece}%
        }
       
\begin{document}

\begin{abstract}

I emphasize analogy between Dp-branes in string theories and solitons in gauge theories 
comparing their common properties and showing differences. 
In string theory we do not have 
the full set of equations which define the theory in all orders of 
coupling  constant as it was in gauge theories, nevertheless  
such solutions have been found  as solutions of low energy 
superstring  effective action carrying the RR charges.
The existence of dynamical RR charged extended objects in string theory 
has been deduced also by considering string theory with mixed boundary 
conditions, when type II closed supersting theory is enriched by 
open strings with Neumann boundary conditions on $p+1$ directions 
and Dirichlet conditions on the remaining 9-p transverse directions.

We will show that for certain excitations of the string/D3-brane 
system  Neumann boundary conditions emerge from 
the Born-Infeld dynamics. 
Here the excitations which are coming down the string with a 
polarization along a direction parallel to the brane are almost completely
reflected just as in the case of all-normal Dirichlet excitations 
considered by Callan and Maldacena,  but now the end of
the string moves freely on the 3-brane  realizing Polchinski's
open string  Neumann boundary condition dynamically. 

In the low energy limit 
$ \omega \rightarrow 0$, i.e. for wavelengths much larger than the string
scale only a small fraction $ \sim \omega^4$ of the energy escapes in
the form of dipole radiation.
The physical interpretation is that a string attached to the 3-brane 
manifests itself as an electric charge, and waves on the string cause
the end point of the string to freely oscillate and produce 
e.m. dipole radiation in the asymptotic outer region. The magnitude 
of emitted power is in fact exactly equal to the                                
one given by Thompson formula  in  electrodynamics.
\end{abstract}

\maketitle

\section{INTRODUCTION}

Probably the best way to introduce D-branes is to recall  
that in non-Abelian gauge field theories there exist  
classical BPS solutions corresponding to new particles,
extended objects,  vortices and monopoles which 
are not present in the perturbative spectrum of the 
quantized theory \cite{no,tp,jz}. In these theories in addition to the 
point particle spectrum of perturbation theory consisting of a
massless photon, massive vector bosons and Xiggs particles there 
also exist stable solutions with masses inversely proportional
to the square of the coupling constant \cite{no,tp,jz}
\be
(Mass)^2_{BPS} = 4\pi M_W /e^2.
\ee 
where $M_W$ is a vector boson mass.
In the case of vortices it is energy per unit of length and 
should be associated with the string tension.  
These states cannot be seen in the weak coupling limit simply 
because the masses of these states become very large when $e \rightarrow 0$,
therefore in gauge theories we have mass hierarchy: the  particle masses are of order 
$O(e^0)$ and the soliton masses are of order $O(1/e^2)$.
These states are invisible in perturbation theory.
One can probably think about these solutions as  " typhoons" of 
gauge bosons. 

The second important message which bring to us these solutions
is that extended objects in gauge theories carry a new type of charge, 
in the given case it is magnetic charge \cite{tp}, which was not present
in perturbative spectrum, as well
\be
g = 2\pi/e.
\ee
The magnetic charge satisfies  the Dirac quantization condition 
\be
e g  = 2\pi n ~~~~~~~~~n  \in  Z,
\ee
which ties together the units of electric and magnetic charges \cite{dir,schw}.
(Generally speaking magnetic charge is realized as a genuine solution without singularities 
when $U(1)$ is embedded into non-Abelian compact group \cite{tp}.)
With the existence of  magnetic charges and also dyons -- particles 
carrying electric and magnetic charges \cite{jz} the theory may exhibit full
electromagnetic duality \cite{mool,wittol,sw}. According to this 
conjecture, the {\it strong coupling} limit of the theory is equivalent
to the {\it weak coupling} limit with ordinary particles and solitons 
exchanged.

In string theory one can also expect solutions which cannot
be seen in the perturbation theory. In string theory we do not have 
the full set of equations which define the theory to all orders of 
coupling  constant as it was in gauge theories, nevertheless  
such solutions have been found  as solutions of low energy 
superstring  effective action \cite{hostr}. The lagrangian
can be represented in the form:
\bea
S = {1 \over 16 \pi G^{(10)}_N } \int d^{10}x \sqrt{-g}   \nonumber\\
\{ e^{-2\varphi} [R + 4(\partial\varphi)^2 - {1 \over 3} H^2]
  - {1 \over 3}G^2 \}
\label{effec}
\eea
where $\varphi$ is the dilaton field, $H$ - the NS-NS gauge field 
and $G$ - RR gauge field have essentially different order in 
dilaton coupling for RR fields. The corresponding equations
have two types of solutions carrying NS-NS or RR charges \cite{hostr}. 

The NS-NS soliton 
is determined by the balance between the first  three terms and 
its tension (tension $\equiv$  mass/volume ), scales as the action
(\ref{effec}) 
\be
(Tension)_{NS-NS} \simeq {1 \over g^{2}_{s}} \label{ten}
\ee 
where $e^{- \varphi} = g_s$ is the string coupling constant. 
In NS-NS sector the electrically 
charged object is the fundamental string of the mass 
$O(g^0_s)$ and is accompanied by the above solitonic 
D5-brane  carring the dual magnetic charge of the mass $O(1/g^2_s)$. 
This is very similar to what we had in gauge theories.

In RR sector the situation is different, all string states
are neutral with respect to RR gauge field and in perturbative 
string spectrum there is no fundamental 
objects carrying elementary  RR charges \cite{rr,pol}. 
In the RR sector the solution of the equations
is determined by the balance between the first
three and the fourth term in the effective action (\ref{effec}) 
and the tension of the solution scales as 
\be
(Tension)_{RR} \simeq {1 \over g_{s}} \label{rrchar}
\ee 
so that both types of RR charges, electric p-brane and 
its dual magnetic $\mbox{6-p}$ brane,  
are solitons of the mass $O(1/g_s)$ which is
smaller than that of the  NS-NS solution (\ref{ten}). 
This behaviour, intermediate between fundamental string 
spectrum $O(g^0_s)$ and solitonic brane $O(1/g^2_s)$, was not 
known in field theory (see Figure 3). 

The existence of dynamical extended objects in string theory 
which carry RR charges can be deduced also by considering 
string theory with mixed boundary conditions \cite{rr}. Indeed 
one can consider type II closed superstring theory  which has 
been completed by adding
{\it  
open strings with Neumann boundary conditions on $p+1$ directions 
and Dirichlet conditions on the remaining 9-p transverse directions} 
\cite{rr,pol,bachas,kiritsis},
\bea
n^{a} \partial_{a}X^{\mu} =0 ,~~~~~~ \mu = 0,...p, \label{nb}\\
X^{\mu} =0 ,~~~~~~ \mu = p+1,...9.  \label{db}
\eea
Here the open string endpoints live on the extended object, D-brane,  
and in this indirect way 
defines a string soliton with $p$ spatial and one timelike coordinates.
The D-branes are surfaces of dimension p on which open string can end. 
This is a consistent string theory if p is even in type IIA
theory and is odd in  type IIB theory \cite{rr}. 
The identification with solitons can be seen if one compute the tension 
\cite{rr,pol}
\be
\tau_{p} = {1\over  g_{s} (2\pi)^{p-1 \over 2} 
(2\pi \alpha^{\prime})^{p+1 \over 2} } \label{dtension}
\ee 
This is the same coupling constant dependence as for the 
branes carrying RR charges (\ref{rrchar})!
The RR charge which carry D-brane is
\be
\mu_{p} = {1\over   (2\pi)^{- {1 \over 2}} 
(4 \pi^2 \alpha^{\prime})^{p-3 \over 2} }. \label{charge}
\ee 
and  satisfies the Dirac quantization condition
\be
\mu_{p} \mu_{6-p} = 2 \pi n ~~~~~~~~n  \in  Z
\ee
with $n=1$. Thus RR gauge fields naturally couple to Dp-branes
and identify Dp-branes as carrier of RR charge \cite{rr}.
The general remark is that the tension of the p-brane should 
have dimension $mass^{p+1}$
and it is indeed so in (\ref{dtension}), but what is important 
is that it is a function of the fundamental string tension 
$1/2\pi \alpha^{\prime}$ (as well as $1/g_s$)
and not just a new parameter of the theory \cite{membra,membra1,gks}. 
The D-branes are indeed the "typhoons" of strings.

\section{THE D-BRANE EXCITATIONS}

The Dp-brane is a dynamical extended object which can fluctuate in 
shape and position and these fluctuations are described 
by the attached open strings. The boundary condition (\ref{db}) allows open 
string ends to be at any place inside the p-brane 
with coordinates $x^0 ,...,x^p$ and $x^{p+1} =...=x^9 =0$.
In this way the open string describes the excitations of the Dp-brane,
and as it is well known the open string 
massless states are vector and  spinor fields 
of ten-dimensional $N=1$ supersymmetric $U(1)$ gauge theory in 
ten dimensions. The 
massless field $A_{\mu}(x^{\nu}), \mu,\nu = 0,...,p$ 
propagates as gauge bosons on the p-brane worldvolume, 
while the other components of the vector potential 
$A_4(x^{\nu}),...,A_9(x^{\nu})$
can be interpreted as oscillation in the position of
the p-brane. Thus the theory on the p-brane worldvolume
is a ten-dimensional supersymmetric $U(1)$ gauge theory
dimensionally reduced to $p+1$ dimensions. The low energy
effective action for these fields is the Dirac-Born-Infeld 
lagrangian \cite{leigh,sch}. The key difference with previously 
known p-branes and supermembranes \cite {duff} is the 
inclusion of a worldvolume electromagnetic field.
  
Callan and Maldacena \cite{cm} showed that the Dirac-Born-Infeld action,
can be used to build a configuration with a semi-infinite 
fundamental string ending on a 3-brane\footnote{ We review the 
favorite case of the 3-brane  first because of its non-singular
behaviour in SUGRA, and secondly it is suggestive of our own world
which is after all 3-dimensional.}, whereby the string is actually
made out of the brane wrapped on $S_2$ (see also \cite{gibbons}). 
The relevant action can be
obtained by computing a simple Born-Infeld determinant, dimensionally
reduced from 10 dimensions
\be
L=-{1\over(2\pi)^3g_s} \int d^4 x \sqrt{1- \vec{E}^2 + (\vec{\partial}x_9)^2},
\ee
where $g_s$ is the string coupling ($\alpha^{\prime} =1$).

The above mentioned theory contains 6 scalars $ x_4,....,x_9~,$ which are 
essentially Kaluza-Klein remnants from the 10-dimensional $N=1$ 
electrodynamics after dimensional reduction to $3+1$ dimensions. As we
explain these extra components of the e.m. field $A_4,...,A_9$ describe
the transverse deviations of the brane $x_4,...,x_9$.

The solution, which satisfies the BPS conditions, is necessarily also       
a solution of the linear theory, the $N=4$ {\it super-}Electrodynamics 
\cite{cm},     
\be                                                                         
\vec{E} = \frac{c}{r^2}\vec{e_r}~,~~~~\vec{\partial}x_9                     
= \frac{c}{r^2}\vec{e_r}~,~~~~x_9 = -\frac{c}{r}~,                          
\label{spike}                                                               
\ee                                                                         
where $c=2 \pi^2 g_s$ is the unit charge,                                       
and can be thought of as setting the distance scale.                        
Here the scalar field represents the geometrical spike,                     
and the electric field insures that the string carries uniform NS           
charge along it. The RR charge of the 3-brane \cite{bfs}
\be
\Omega^{MNL} = \epsilon^{abc}\partial_a X^{M} 
\partial_b X^{N} \partial_c X^{L}
\ee
cancels out on the            
string, or rather the tube behaves as a kind of RR dipole whose             
magnitude can be ignored when the tube becomes thin.  It was demonstrated            
in \cite{cm} that the infinite electrostatic energy of the point       
charge can be reinterpreted as being due to the infinite length of the      
attached string. The energy per unit length comes from the electric field   
and corresponds exactly to the fundamental string tension.                  

Polchinski, when he introduced D-branes as objects on which strings can
end, required that the string has Dirichlet (fixed) boundary conditions
for coordinates normal to the brane (\ref{db}), and   Neumann (free) boundary 
conditions for coordinate directions parallel to the brane (\ref{nb})
\cite{rr,pol,leigh}. 

It was
shown in \cite{cm} that small fluctuations which are normal to both the 
string and the brane are mostly reflected back with a 
$phase~ shift  \rightarrow \pi$
which indeed corresponds to Dirichlet boundary condition (\ref{db}).
See also \cite{larus} and \cite{rey} for 
a supergravity treatment of this problem.

In the paper \cite{konstantin} it was demonstrated that P-wave 
excitations which are coming down the string with a
polarization along a direction parallel to the brane are almost completely
reflected just as in the case of all-normal excitations, but the end of
the string moves freely on the 3-brane, thus realizing Polchinski's
open string    Neumann boundary condition (\ref{nb}) dynamically. 
As we will see a superposition of excitations of the 
scalar $x_9$ and of the electromagnetic field 
reproduces the required behaviour,
e.g. reflection of the geometrical fluctuation with a 
$phase~ shift  \rightarrow 0$
(Neumann boundary condition).

In \cite{konstantin} it was also observed 
the electromagnetic dipole radiation which escapes to infinity
from the place where the string is attached to the 3-brane. 
It was shown that in the low energy limit
$ \omega \rightarrow 0$, i.e. for wavelengths much larger than the string
scale a small fraction $ \sim \omega^4$ of the energy escapes to infinity in
the form of electromagnetic dipole radiation.
The physical interpretation is that a string attached to the 3-brane
manifests itself as an electric charge, and waves on the string cause
the end point of the string to freely oscillate and produce
electromagnetic dipole radiation in the asymptotic outer region of the 3-brane.
This result is also in a good agreement with the interpretation of 
the open string ending on D-brane  (\ref{nb}),
according to which the open string states 
describe the excitations of the D-brane. 
Indeed as we will see open string massless vector boson 
can propagate  inside the D-brane and 
in our case this excitation is identified 
with electromagnetic dipole radiation.
Thus not only in the static case, but also in a more general dynamical 
situation the above interpretation remains valid.
This result provides additional support to the idea that 
the electron (quark)  may be understood as the end of a fundamental string 
ending on a D-brane.

\section{THE LAGRANGIAN AND THE E\-QUA\-TI\-ONS}

Let us write out the full Lagrangian which contains both 
electric and magnetic fields, plus the scalar $x_9 \equiv \phi$
\cite{konstantin}
\bea
L=-\int d^4x \sqrt{Det}~\\
\nonumber
where~~~Det=1+ \vec{B}^2 - \vec{E}^2 -(\vec{E} \cdot \vec{B})^2 \\ \nonumber-
   (\partial_0 \phi)^2(1+\vec{B}^2) +(\vec{\partial}\phi)^2
   +(\vec{B}\cdot \vec{\partial} \phi)^2 \\ \nonumber
   - (\vec{E}\times \vec{\partial}\phi)^2+
   2\partial_0 \phi(\vec{B}[\vec{\partial} \phi\times\vec{E}]) 
\label{lagrfull}
\eea
We will proceed by adding a fluctuation to the background values 
(\ref{spike}) : 
$$ \vec{E}=\vec{E}_0 + \delta \vec{E},~~\vec{B}= 
\delta \vec{B},~~\phi=\phi_0 +\eta~.$$
Then keeping only terms in the $Det$  which are linear and quadratic
in the fluctuation we will get 
\bea
\delta Det =  \delta\vec{B}^2 - \delta\vec{E}^2 - 
(\vec{E_{0}} \delta\vec{B})^2  - (\partial_0 \eta)^2 \\ \nonumber
+ (\vec{\partial}\eta)^2 
+ (\delta\vec{B} \vec{\partial} \phi)^2
- (\vec{E_0} \times \vec{\partial}\eta)^2 
- (\delta\vec{E} \times \vec{\partial} \phi)^2 \\ \nonumber
-2(\vec{E_0} \times \vec{\partial}\eta)
(\delta\vec{E} \times \vec{\partial}\phi)
-2(\vec{E_0} \delta\vec{E}) + 2 (\vec{\partial}\phi \vec{\partial}\eta)
\eea
Note that one should keep the last two linear terms because they produce 
additional quadratic terms after taking the square root.
These terms involve the longitudinal polarization of the 
e.m. field and cancel out at quadratic order.
The resulting quadratic Lagrangian is \cite{konstantin}
\bea
2L_q = \delta\vec{E}^2(1+(\vec{\partial} \phi)^2) - \delta\vec{B}^2 
+ (\partial_0\eta)^2 \\ \nonumber -
(\vec{\partial}\eta)^2(1-\vec{E_0}^2) 
+ \vec{E_0}^2 
( \vec{\partial}\eta \cdot \delta\vec{E})~.
\label{full}
\eea
Let us introduce the gauge potential for the fluctuation part of the e.m.
field as $(A_0,\vec{A})$ and substitute the values of the background 
fields from (\ref{spike})
\bea
2L_q = (\partial_0 \vec{A}- \vec{\partial}A_0)^2 (1+{1\over r^4})-
       (\vec{\nabla}\times\vec{A})^2  \\ \nonumber
       + (\partial_0\eta)^2
       -(\vec{\partial}\eta)^2(1-{1\over r^4}) + 
       {1\over r^4} (\partial_0 \vec{A}- \vec{\partial}A_0)
       \cdot \vec{\partial}\eta ~.
\label{fluctlag}
\eea
The equations that follow from this lagrangian contain dynamical 
equations for the vector potential and for the scalar field, 
and a separate equation which represents a constraint. These
equations in the Lorenz gauge
$\vec{\partial}~\vec{A}=\partial_0 A_0$ are
\bea
 \label{alpha}
-\partial_0^2 \vec{A}(1+{1\over r^4}) + \Delta\vec{A}+
{1\over r^4}\vec{\partial}\partial_0(A_0+\eta) = 0 \\
\label{beta}
-\partial_0^2 A_0 + \Delta A_0 + 
\vec{\partial} {1\over r^4} \vec{\partial}(A_0+\eta) - 
\vec{\partial} {1\over r^4} \partial_0\vec{A} = 0  \\ 
\label{gamma}
-\partial_0^2 \eta~ + \Delta \eta~~ - 
\vec{\partial} {1\over r^4} \vec{\partial}(A_0+\eta) + 
\vec{\partial} {1\over r^4} \partial_0\vec{A} = 0
\eea
Equation (\ref{beta}) is a constraint: the time derivative of
the {\it lhs} is zero, as can be shown using the equation of motion 
(\ref{alpha}).

Let us choose $A_0=-\eta$. This condition can be viewed as (an attempt to)
preserve the BPS relation which holds for the background: 
$\vec{E}=\vec{\partial } \phi$. Another point of view is that this fixes
the general coordinate invariance which is inherent in the Born-Infeld 
lagrangian in such a way as to make the given perturbation to be
normal to the surface. Of course transversality is insured automatically
but this choice makes it explicite. The general treatment of this 
subject can be found in \cite{sch}.

With this condition the equations (\ref{beta}) and 
(\ref{gamma}) become the same, and
the first equation is also simplified \cite{konstantin}:
\bea
-\partial_0^2 \vec{A}(1+{1\over r^4}) + ~\Delta\vec{A} = 0~, \\
-\partial_0^2 \eta + \Delta \eta +
\vec{\partial} {1\over r^4} \partial_0\vec{A} = 0~.
\label{seqns}
\eea
This should be understood to imply that once we obtain a solution,
$A_0$ is determined from $\eta$, but in addition we are now obliged to 
respect the gauge condition which goes over to 
$\vec{\partial}\vec{A}=-\partial_0 \eta$.

\section{NEUMANN BOUNDARY CON\-DI\-TI\-ONS FROM BORN-INFELD DYNAMICS
 \cite{konstantin}  } 
We will seek a solution with definite energy                                                 
(frequency $w$) in the following form: $\vec{A}$ should have                                 
only one component $A_z$, and $\eta$ be an $l=1$ spherical $P$-wave                          
$$                                                                                           
A_z= \zeta(r)~e^{-i\omega t}~~,~~~\eta={z\over r}~\psi(r) ~e^{-i\omega t}                    
$$                                                                                           
The geometrical meaning of such a choice for $\eta$ is explained in Figure 1,                
and the particular choice of $z$ dependence corresponds to the polarization                  
of the oscillations along the $z$ direction of the brane.                                    
With this ansatz the equations become                                                        
\bea                                                                                         
\label{oldfriend}                                                                            
\left(1+{c^2\over r^4}\right)~\omega^2 \zeta +                                               
{c^2\over r^2}\partial_r(r^2 \partial_r \zeta)=0~,\\ \nonumber                                        
{z\over r}~ \omega^2 \psi +                                                                  
{z\over r}~ {1\over r^2}\partial_r(r^2 \partial_r \psi) +                                    
{z\over r}~ {2\over r^2}~\psi -\\
 i \omega \partial_z\left({c^2\over r^4}\zeta\right) = 0~,    
\eea                                                                                         
with the gauge condition becoming $\partial_r\zeta =i \omega \psi$.                          
It can be seen again, that the second equation follows from the first by                     
differentiation. This is because the former coincides with the constraint                    
in our ansatz.                                                                               
                                                                                             
\begin{figure}
\centerline{\hbox{\psfig{figure=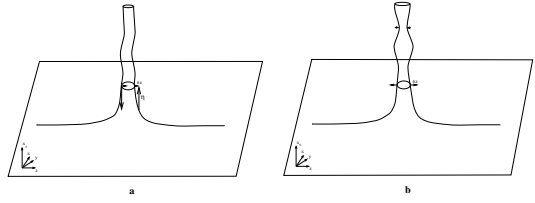,width=7cm}}}
\caption[fig1]{ In order that all points on the $S_2$ section of the
tube ( which is schematically shown here as a circle) move all in
the same direction $\hat z$ by an equal distance $\delta z$, the field 
$\eta$ has to take on different values at, say, opposite points of the 
$S_2$. In effect, 
$ \eta=\delta{-1\over r}={1 \over r^2}{z \over r}\delta z$. 
If, on 
the other hand, it is taken to be an $S$-wave (as in the paper \cite{rey})
that would correspond to 
Fig 1b, which at best is a problematic `internal' degree of freedom of the
tube.
} 
\label{fig1}
\end{figure}

Therefore  the problem is reduced to finding the solution of a                           
single scalar equation, and determining the other fields through                         
subsidiary conditions. The equation (\ref{oldfriend})                                    
itself surprisingly turned out to be                                                     
the one familiar  from \cite{cm} for the transverse fluctuations.                        
There it was solved by going over to a coordinate $\xi$ which                            
measures the distance radially along the surface of the brane                            
$$                                                                                       
\xi(r) = \omega \int\limits_{\sqrt{c}}^{r} du \sqrt{1 +{c^2 \over u^4}}~~,                
$$                                                                                       
and a new wavefunction                                                                   
$$                                                                                       
\tilde{\zeta}=\zeta~(c^2+ r^4)^{1/4}~.                                                   
$$                                                                                       
This coordinate behaves as $\xi \sim r$ in the outer region                              
($r\rightarrow\infty$)                                                                   
and $\xi \sim -1/r$ on the string ($r\rightarrow 0$).                                    
The exact symmetry of the equation                                                       
$r \leftrightarrow c/r$ goes over to $ \xi \leftrightarrow - \xi$.                       
The equation, when written in this coordinate becomes just the free                      
wave equation, plus a narrow symmetric potential at $\xi \sim 0$                         
\be                                                                                      
-{d^2\over d\xi^2}\tilde{\zeta}+                                                         
\frac{5 c^2 /\omega^2}{(r^2+c^2/r^2)^3}\tilde{\zeta}=\tilde{\zeta}~.                     
\ee                                                                                      
The asymptotic wave functions can be constructed as plain waves in $\xi$,                
$$                                                                                       
\zeta(r) = (c^2 + r^4)^{-1/4} e^{\pm i \xi(r)}~,                                         
$$                                                                                       
or in the various limits:                                                                
\bea                                                                                     
\nonumber                                                                                
r\rightarrow ~0~~~~\zeta \sim ~~{1 \over \sqrt{c}} e^{\pm i \xi(r)}~,\\                                     
\nonumber                                                                                
r\rightarrow \infty~~~~\zeta \sim {1 \over r}e^{\pm i \xi(r)}~.                          
\eea                                                                                     
These formulae give us the asymptotic wave function in the regions                       
$ \xi \rightarrow \pm \infty$,                                                           
while around $\xi = 0~(r=1)$ there is a symmetric repulsive potential                    
which drops very fast $ \sim 1/\xi^6$ on either side of the origin.                      
The scattering is described by a single dimensionless parameter                          
$\omega \sqrt{c}$, and in the limit of small $\omega$ and/or coupling                    
$c=2 \pi^2 g_s$ the potential becomes narrow and high, and  can be                           
replaced by a $\delta$-function with an equivalent  area                                 
$\sim {1\over \omega \sqrt{g_s}}$ under the curve.                                       
Therefore the scattering matrix becomes almost                                           
independent of the exact form of the potential.                                          
We refer the reader to pp 18-20 of the thesis \cite{thesis}                              
for the more detailed calculation of transmission amplitude (\ref{transition}).                            
The end result is that                                                                   
most of the    $\zeta$ amplitude is reflected back with a phase shift close to $\pi$.
                                                                                         
In order to obtain $\psi$ (and $\eta$) we need to differentiate                          
$\zeta$ with respect to $r$:                                                             
\bea                                                                                      
i\omega\psi= {-1\over4} {4r^3\over(c^2+r^4)^{5/4}}e^{\pm i \xi(r)} \nonumber\\                      
\pm{i\omega\over(c^2+r^4)^{1/4}}                                                         
\left(1 +{c^2 \over r^4}\right)^{1/2} e^{\pm i \xi(r)}~~.                                
\eea                                                                                      
Again it is easy to obtain the simplified limiting form:                                                                           
\bea                                                                                                                                                                     
r\rightarrow 0~~~~~i \omega \psi                                                         
\sim \left(-{r^3 \over c^{5/2} } \pm {i\omega \sqrt{c}\over r^2}\right) 
e^{\pm i \xi(r)}\nonumber\\                                                                                
r\rightarrow \infty~~~~i \omega \psi                                                     
\sim \left({-1 \over r^2} \pm {i\omega \over r}\right) e^{\pm i \xi(r)}  \nonumber                
\eea                                                                                     
This                                                                                     
brings about several consequences for $\psi$. Firstly, it causes                         
$\psi$ to grow as $\sim 1/r^2$ when $r\rightarrow 0$. This is the                        
correct behaviour because when converted to displacement in the                          
$z$ direction, it means constant amplitude \cite{konstantin}:                                              
\be                                                                                      
 \eta={z\over r}\psi=\delta{-c\over r}={z \over r}{c \over r^2}\delta z~~%
\Rightarrow ~~\delta z \sim const~~.                                                     
\ee                                                                                      
Secondly, the $i$ that enters causes the superposition of the                            
incoming and reflected waves to become a cosine from a sine, as is                       
the case for $\zeta$ waves. {\it This corresponds to a $0$ phase shift 
for the $\eta$ wave (Figure 2) and  implies  therefore  Neumann boundary condition ,
that is it allows open string ends to be at any place inside the p-brane . }

\section{ELECTROMAGNETIC~DIPO\-LE RA\-DIA\-TION~OF~OSCILLATING
D- BRANE \cite{konstantin} } 

Because of the $\omega$ factor in the gauge condition we                                           
need to be careful about normalizations, thus we shall choose to                                   
fix the amplitude of the $\delta z$ (or $\eta$) wave to be independent of                          
frequency $\omega$.                                                                                
Then the magnitude of the electromagnetic field in the inner region becomes                        
independent of $\omega$ as well, thus the amplitude of $\zeta$ wave at 
$r\rightarrow 0$ is 
$$
A_z = \zeta_0 e^{-i \omega ( t + c/r) }
$$
Combined with the transmission factor                                             
\be
T=-i \omega\sqrt{c} + O(\omega^2), \label{transition}
\ee 
this gives the following form of the                                             
dipole radiation field at infinity\footnote{The unit of electric charge in our notation            
is $2 \pi^2 g_s=e^2$. This is because of the scaling of the fields needed                              
to get the U(1) action with $1/e^2$ in front of it.}                                               
\bea                                                                                                
A_z =  {T \cdot \zeta_0 \over r }e^{-i \omega t} = 
{-i\omega \sqrt{2 \pi^2 g_s} \over r } \zeta_0 e^{-i \omega t}\nonumber\\=                                 
{-i \omega e \zeta_0\over r } e^{-i \omega t}={\dot{d} \over r },                                      
\label{dipolerad}                                                                                  
\eea 
where
$$
d = e \zeta_0 e^{-i \omega t}.
$$                                                                                               
In order to make a comparison with Thompson formula                                                 
$$I={4\over 3}\omega^4 e^2 {\zeta_0}^2
$$ for the total                                                       
power emitted by an oscillating dipole, we note that the agreement                                 
is guaranteed if the exact coefficient  behind $-i \omega\sqrt{c}$                                                             
in (\ref{transition}) is equal to one. As it was shown                                                     
in \cite{thesis}, the approximate computation of the transmission amplitude                        
$T$  (\ref{transition})  gives $T \sim  -i .92 \omega \sqrt{c}$. 
The later computation of this coefficient using the numerical 
solution of the equation (\ref{oldfriend}) shows that 
it is indeed equal to one. To solve properly the equation on
a computer it is necessary to introduce new variables \cite{kons}
\be
y = r- {c \over r},~~or~~ r(y) = {y+\sqrt{y^2 +  4c} \over 2},
\ee
so that the equation (\ref{oldfriend}) take the form
\be
\omega^2 (c^2 + r(y)^4) \zeta + (c + r(y)^2) \partial_y (c + r(y)^2) \partial_y \zeta =0
\ee
and boundary condition at $r_0 = \omega c/ 2\pi n$, $n \rightarrow \infty$ the form
\bea                                                                                                                                                                     
\zeta(y_0) = {1 \over \sqrt{c}},~~\zeta^{\prime}(y_0) = 
-i {(2\pi n)^2 \over \omega c^{3/2}}, \nonumber \\  
at~~~~y_0 = {\omega c \over 2\pi n}- {2\pi n \over \omega}    \rightarrow -\infty .\nonumber            
\eea  
Measuring the amplitudes $y\zeta(y) \pm (y + {\pi \over 2\omega}) 
\zeta(y + {\pi \over 2\omega})$
at $y \rightarrow \infty (r \rightarrow \infty)$ one can get that the 
coefficient  behind $-i \omega\sqrt{c}$ 
in (\ref{transition}) is equal to one.
Thus the magnitude of emitted power is in fact exactly equal to the                                
one given by Thompson formula  in electrodynamics\footnote{see also 
\cite{callan,costa,bachas1,janik,rho}}.

\begin{figure}
\centerline{\hbox{\psfig{figure=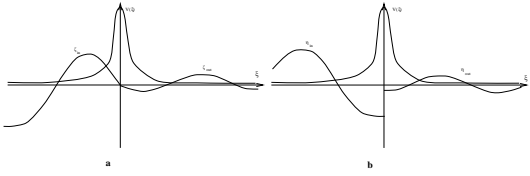,width=7cm}}}
\caption[fig2]{ The figure 2a depicts the scattering of the 
$\zeta$ wave. Note the discontinuity in the derivative which is 
proportional to $1/\omega$. Figure 2b shows the scattering of
the $\eta$ wave. Being the derivative of $\zeta$, it 
results in a discontinuity of the function itself,
making it into a cosine, which means free (Neumann) 
boundary condition at $\xi =0$.
} 
\label{fig2}
\end{figure}

In conclusion, we need to analyze the outgoing scalar wave. This wave has              
both real and imaginary parts, the imaginary part is $\sim 1/r^2$             
and drops faster than radiation. The real part                                       
does contribute to the radiation at spatial infinity,                                  
as can be shown from the integral of the energy density                                
$\int (\partial_r \eta)^2 d^3r \sim                                                    
\int \omega^4/r^2 \cdot 4\pi r^2 dr \sim \omega^4$.                                    
This is not altogether surprising, as we are dealing with a                            
supersymmetric theory where the different fields are tied together.                    
In fact, it is easy to show that  the fields                                           
still do asymptotically satisfy the BPS conditions at infinity, and                    
$1/4$ of the SUSY is preserved.                                                        
Thus the observer at spatial infinity will see both an                                 
electromagnetic dipole radiation field and a scalar wave.                              
Interestingly, the direction dependences of the two conspire                           
to produce a spherically symmetric distribution                                        
of the energy radiated.                                                              

\begin{figure}
\centerline{\hbox{\psfig{figure=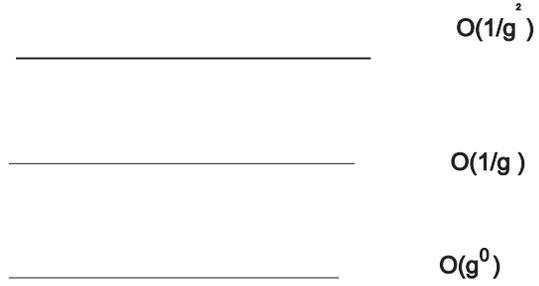,width=7cm}}}
\caption[fig3]{ In  NS-NS sector the electrically 
charged object is the fundamental string of the mass 
$O(g^0_s)$ and is accompanied by the solitonic 
D5-brane  carring the dual magnetic charge of the mass $O(1/g^2_s)$. 
In the RR sector both types of RR charges, electric p-brane and 
its dual magnetic $6-p$ brane,  
are solitons of the mass $O(1/g_s)$ which is
smaller than that of the  NS-NS solution. 
This behaviour, intermediate between fundamental string 
spectrum $O(g^0_s)$ and solitonic brane $O(1/g^2_s)$, was not 
known in field theory.
} 
\label{fig3}
\end{figure}

I wish to thank the organizers of the conference QG99 in Villasimius and 
especially Vittorio de Alfaro for the invitation and for 
arranging interesting and stimulating meeting. I also thank Konstantin 
Savvidy for the enjoyable collaboration on the work reported here.
The work was supported in part by the EEC Grant no. HPRN-CT-1999-00161 
and ESF Grant "Geometry and Disorder: From Membranes to Quantum Gravity".

\end{document}